\documentclass[aps,preprint,superscriptaddress]{revtex4}
\usepackage[dvips]{graphicx}
\usepackage{epsfig,amsmath,amsfonts,amssymb,wasysym,color}

\begin{document}
\def\be{\begin{equation}}
\def\ee{\end{equation}}
\def\bfi{\begin{figure}}
\def\efi{\end{figure}}
\def\bea{\begin{eqnarray}}
\def\eea{\end{eqnarray}}

\newcommand{\reff}[1]{(\ref{#1})}
\def\x{{\bf x}}
\def\y{{\bf y}}
\def\k{{\bf k}}
\def\r{{\bf r}}
\def\op{{\varphi}}
\def\H{{\mathcal H}}
\def\tm{{\tau_{\rm micr}}}
\def\rmd{{\rm d}}
\def\gl{{\rm (g)}}
\def\lo{{\rm (l)}}
\def\lsi{{\rm (LSI)}}
\def\ft{{\rm (FT)}}
\def\mF{{\mathcal F}}
\def\mA{{\mathcal A}}

\title{%
The non-equilibrium response of the critical Ising model:
Universal scaling properties and Local Scale Invariance}

\author{Federico Corberi}
\email{corberi@sa.infn.it}
\affiliation{\small Dipartimento di Matematica ed Informatica, Universit\`a
di Salerno, via Ponte don Melillo, 84084 Fisciano (SA), Italy.}
\author{Andrea Gambassi}
\email{gambassi@mf.mpg.de}
\affiliation{\small Max-Planck-Institut f\"ur Metallforschung,
Heisenbergstr.~3, D-70569 Stuttgart, Germany.}
\affiliation{\small Institut f\"ur Theoretische und Angewandte Physik,
Universit\"at Stuttgart, Pfaffenwaldring~57, D-70569 Stuttgart, Germany.}
\author{Eugenio Lippiello}
\email{lippiello@sa.infn.it}
\affiliation{\small Dipartimento di Scienze Fisiche, Universit\`a
di Napoli ``Federico II'', 80125 Napoli, Italy.}
\author{Marco Zannetti}
\email{zannetti@sa.infn.it}
\affiliation{\small Dipartimento di Matematica ed Informatica,
Universit\`a di Salerno, via Ponte don Melillo, 84084 Fisciano (SA), Italy.}

\begin{abstract}
Motivated by recent numerical findings [M.~Henkel, T.~Enss, and
M.~Pleimling, J.~Phys.~A: Math. Gen.~{\bf 39} (2006) L589] we re-examine
via Monte Carlo simulations the linear response function
of the two-dimensional Ising model with Glauber dynamics
quenched to the critical point. 
At variance with the results of Henkel {\it et al.}, 
we detect discrepancies between the actual scaling behavior
of the response function and the prediction of Local Scale
Invariance. 
Such differences are clearly visible in the 
{\it impulse} autoresponse function, whereas they are
drastically reduced in 
{\it integrated} response functions.
Accordingly, the scaling form predicted on
the basis of Local Scale Invariance simply provides an 
accurate fitting form for some quantities but cannot be 
considered to be exact.
\end{abstract}

\maketitle

\section{Introduction} \label{intro}

The non-equilibrium collective relaxation of pure systems quenched at
or below their critical points has been recently the subject of a
renewed interest in connection with the fact that two-time quantities
--- such as response and correlation functions --- display a scaling
behavior similar to the one observed in glassy
systems~\cite{Cugliandolo,God-rev,cg-rev}. In addition, at the critical
point, this scaling behavior is characterized by a certain degree of
{\it universality}~\cite{God-rev,cg-rev} which renders it largely independent
of the microscopic details of the systems and allows its
investigation via suitable minimal models, either on the lattice or in the
continuum. 
The former can be easily studied via Monte Carlo simulations
(or, in a limited number of cases, exact solutions are available), whereas
powerful field-theoretical methods have been developed for the
latter. 
       
In what follows we will be concerned with the behavior of the
two-time response function. 
Consider a lattice model characterized by degrees of freedom 
$\sigma_\x$ with $\x \in {\mathbb Z}^d$, interacting via an Hamiltonian
$\H[\{\sigma_\x\};\{h_\x(t)\}] = \H_0[\{\sigma_\x\}] - \sum_\x \sigma_\x h_\x(t)$,
where $h_\x(t)$ is a time-dependent external field. In the case of the Ising model
$\sigma_\x = \pm 1$, 
$\H_0[\{\sigma_\x\}] = - J \sum_{\langle\x\y \rangle}
\sigma_\x\sigma_\y$ ($J>0$), 
the sum being over pairs $\langle\x\y \rangle$ of nearest neighboring sites. 
The stochastic 
dynamics of the model, due to the coupling to a thermal bath of
temperature $T$, 
can be implemented according to different rules 
and yields a time dependence of the value of the local degree of
freedom $\sigma_\x(t)$.  The linear two-time 
response function is defined as
\be
R_{r}(t,s) = \left.\frac{\delta \langle \sigma_\x(t)\rangle}{\delta
h_\y(s)}\right|_{h=0}
\label{resp}
\ee
where $r=|\x -\y|$,  $\langle\ldots\rangle$ indicates the average 
over the possible realizations of the stochastic dynamics, 
and invariance under space translations has been assumed. 
Causality implies that $R_r(t,s)$
vanishes for $s>t$ and therefore we shall assume $t>s$ in what
follows. 
In addition, we shall primarily consider the 
local response function (also referred to as {\it
auto}response function)
\be
R(t,s) \equiv  R_{r=0}(t,s).
\label{autoresp}
\ee
If the system is in thermal equilibrium with the bath, the dynamics is
invariant under time translations 
and therefore $R_r(t,s)$ actually depends only on $\tau \equiv t-s$, i.e.,
$R_r(t,s)=R_r^{\rm (eq)}(\tau;T)$.
Time-translational invariance is naturally broken if, at 
time $t=0$, the temperature of the bath 
is suddenly changed from an initial value $T_{\rm i}$ --- which we shall
assume to be high enough to yield a disordered state in the
system %
--- to a final value $T_{\rm f} \neq T_{\rm i}$. 
After the quench, the system undergoes a
non-equilibrium relaxation towards 
the new equilibrium state at the temperature $T_{\rm f}$, which is
attained after a time $t_{\rm eq}$.
In some circumstances, however, $t_{\rm eq} = \infty$ and 
the dynamics retains its non-equilibrium character forever.
Interestingly enough, a robust scaling behavior emerge in this regime 
as it is revealed by
two-time quantities such as
the linear response function~\reff{resp}.
In the paradigmatic case of ferromagnetic systems,  $t_{\rm eq} =
\infty$ 
--- in the thermodynamic limit ---
if the quench occurs at $T_{\rm f}\le T_c$,
where $T_c$ is the critical temperature of the system. 
In such a case, the behavior of $R_r(t,s)$ depends on the relation
between $t$ and $s$. 
We restrict our analysis to the case $s,\tau \gg \tm$,
where $\tm$ is a microscopic
time scale set by the dynamics and such that for $s,\tau < \tm$ 
non-universal corrections to scaling appear.
In the sector $s \gg \tau \gg \tm$,
referred to as {\it short-time separation regime} (STSR),
time-translational invariance 
and time-reversal symmetry 
are recovered in local quantities,
resulting in an equilibrium-like dynamics:
\be
R(t,s)=R^{\rm (eq)}(\tau; T_{\rm f}).
\label{1}
\ee
This {\it quasi-equilibrium} behavior is quite generally 
expected in systems with slow dynamics (see, e.g.,
Refs.~\cite{Cugliandolo,God-rev,cg-rev}). 
By contrast, for  $s\gg\tm$ and fixed 
$u \equiv t/s > 1$, referred to as 
{\it aging regime}, one expects the scaling behavior
\be
R(t,s)=R^{\rm (ag)}(t,s)=s^{-1-a}F_R(u)\;,
\label{genscalag}
\ee
which is characterized by the scaling function $F_R$ and the exponent
$a$. 
These being the behaviors in two different regimes, 
it remains to be clarified:
(a) how the crossover between the forms~\reff{1} and~\reff{genscalag} 
actually occurs in $R(t,s)$ and (b) how $F_R(u)$ behaves.

For quenches at the critical temperature these issues can be addressed
via scaling arguments~\cite{God2,God-rev} and the 
renormalization-group (RG)
approach~\cite{JSS}, which yield
\be
R(t,s) = A_R s^{-1-a} u^\theta (u-1)^{-1-a}f_R(u)   
\label{2}
\ee
where the exponent $a$ is related to the usual static and dynamic critical 
exponents $\eta$ and $z$ by 
\be
\label{expa}
a=\frac{d-2+\eta}{z},
\ee
$\theta $ is the so-called
initial-slip exponent~\cite{JSS}, $A_R$ is a
{\it non-universal} constant (fixed by the condition 
$f_R(u\gg 1) = 1$) and $f_R(u)$ is a {\it universal} scaling function
such that~\cite{nota1} 
\be
f_R(1^+)\neq 0 \quad \mbox{and}\quad \mbox{{\it finite}}. 
\label{aeqa'}
\ee
$f_R(u)$ 
can be calculated by means of a variety of field-theoretical techniques.
As a consequence of Eq.~\reff{aeqa'}, 
Eq.~\reff{2} can be cast in the form
\be
R(t,s)=R^{\rm (eq)}(\tau;T_c)R^{\rm (off)}(u)
\label{3}
\ee
where
\be
R^{\rm (eq)}(\tau;T_c) = B_R \tau^{-1-a}
\label{3a}
\ee
is the {\it equilibrium} critical response function with $B_R=A_Rf_R(1^+)$ and
\be
R^{\rm (off)}(u) = u^\theta f_R(u)/f_R(1^+)
\label{3b}
\ee 
is the non-equilibrium contribution.
In the STSR $R^{\rm (off)}(u)\simeq 1$ and the response function~\reff{2} 
reduces to the form~\reff{3a}, in agreement with what expected from
Eq.~\reff{1}. In the aging regime $R^{\rm (off)}(u)\neq 1$ and
the invariance under 
time-translations, typical of equilibrium, 
is broken. Equation~\reff{3} shows that in the
quench at $T_{\rm f}=T_c$ 
the crossover between the quasi-equilibrium $R^{\rm (eq)}$ and 
the aging $R^{\rm (ag)}$ behavior occurs multiplicatively via the
factor $R^{\rm (off)}$.

Unfortunately, no equivalent powerful methods are
available for the analytical investigation of the behavior of the
response function after a quench to $T_{\rm f} < T_c$. 
Nonetheless, the available 
numerical simulations of the Ising model~\cite{noiTRM,noiRd2,noi},
the exact results for the Glauber-Ising  
chain~\cite{God1,Lippiello} quenched to 
$T_{\rm f}=0$~\cite{footnote1d}, 
and the predictions of 
approximate theories of phase-ordering~\cite{berthier,EPJ,Mazenko} 
suggest that the response function for 
scalar systems in the {\it aging regime}   
can be cast in the form of Eq.~\reff{2}
where $a$ is a scaling exponent whose behavior has been 
investigated in Refs.~\cite{noiTRM,noia}, with
\be
\label{fRPO}
f_R(u) = (u-1)^{1/z_g} \;.
\ee
(See, however, Refs.~\cite{MHenkel,Janke} where $f_R(u) \equiv 1$ has
  been numerically inferred by looking at the so-called 
thermoremanent magnetization --- 
$\chi (t,[0,s])$ and $\hat \chi_{k=0} (t,[0,s])$, 
see Eqs.~\reff{intreal} and~\reff{intmom} below --- for 
the Ising, $3$- and $8$-state Potts models.)
Here $z_g$ is the growth exponent which controls the time-evolution of
the typical domain size $L(t)\sim t^{1/z_g}$ in systems with discrete
symmetries (e.g., $z_g=2$ for the Ising model with 
Glauber dynamics)~\cite{Bray}.
Note that, at variance with the prediction of the scaling and RG
analysis for the quench to $T_{\rm f}=T_c$, 
$f_R(u)$ {\it vanishes} for $u\rightarrow 1^+$. 
Let us stress that, while for $T_{\rm f} =T_c$ scaling arguments and RG
approach predict Eq.~\reff{2} to be obeyed for {\it every} $t$ and
$s$, for $T_{\rm f}<T_c$ this expression holds {\it only} in the {\it
aging regime}.

Due to the lack of general predictions for the scaling function $f_R(u)$ of the
response function it is particularly interesting the proposal  by Henkel et
al.~--- referred to as local scale invariance
(LSI)~\cite{Henkel,H-new,H-old} ---  stating that the scaling form  of
the response function is constrained by requiring its
covariance under a group of {\it local} scale transformations
which generalize the {\it global}  transformations ($b>0$) $\x \mapsto
b \x$, $(t,s) \mapsto (b^z t,\, b^z s)$ underlying the scaling
behavior of critical dynamics.

In this paper we investigate the extent to which LSI describes
correctly the scaling behavior of the response function $R(t,s)$,
focusing  on the  two-dimensional Ising model with Glauber dynamics
after a quench from the disordered state {\it to the critical point}.
In particular we revisit the recent findings of
Ref.~\cite{HenkelLSI1}.

The presentation is organized as follows:
In Sec.~\ref{sec-LSI} we summarize the predictions of the different
versions of LSI, highlighting their qualitative features and
discussing the comparison with the available analytical and numerical
results for a variety of different systems undergoing non-equilibrium
relaxation after a quench at $T_{\rm f}\le T_c$. In particular we
discuss those features of LSI that we shall specifically investigate in the
numerical analysis.
In Sec.~\ref{MC} we present the 
results of our Monte Carlo simulations for quenches to $T_{\rm f}=T_c$. 
After the description of the numerical procedure,
we discuss the data for the impulse response function
$R(t,s)$ in Subsec.~\ref{impulsive} and for the associated local
and global  integrated response functions $\chi_I^\lo(t,s)$
[cf. Eq.~\reff{chiIl-def}] and $\chi_I^\gl(t,s)$ [cf. Eq.~\reff{chiIg-def}] 
in Subsec.~\ref{integrated-l} and~\ref{integrated-g},
respectively, comparing them with previous results and
with the predictions of LSI.
In Sec.~\ref{quenchsotto} we briefly 
present the comparison between what is currently known about
the response function $R(t,s)$, after quenches to $T_{\rm f} <T_c$, and
LSI. 
We point out that the aging part of $R(t,s)$ is actually described by
a recently proposed version of LSI.  
A final summary and our conclusions are presented in Sec.~\ref{conclusions}.

\section{Local Scale Invariance: Scaling forms for the response
function}

\label{sec-LSI}

For the impulse autoresponse function $R(t,s)$, 
LSI in its original
form~\cite{H-old,Henkel} (which we refer to as LSI.0) predicts ($u = t/s>1$)
\be
R(t,s)=A_R s^{-1-\alpha} u^\beta(u-1)^{-1-\alpha}
\label{lsi0}
\ee
and
\be
R_r(t,s)= R(t,s)\Phi(r(t-s)^{-1/z}),
\label{lsispace}
\ee 
where the exponents $\alpha$, $\beta$ and the amplitude $A_R$ are
undetermined whereas $\Phi$ is defined in terms of
special functions in Ref.~\cite{H-old}. 
These predictions are supposedly quite general and 
they apply to a variety of systems undergoing 
non-equilibrium relaxation, either due to critical ($T_{\rm
f} = T_c$) or phase ordering ($T_{\rm f} <  T_c$) dynamics. 
As in the case of dynamics at the critical point 
(see Eqs.~\reff{2} and~\reff{aeqa'}), the structure of
Eq.~\reff{lsi0} is such that time-translational invariance
is recovered in the STSR: $R(t=s+\tau,s\gg\tau) = A_R
\tau^{-1-\alpha}$ and 
this stationary part crosses over to the 
non-equilibrium behavior via a multiplicative factor as in
Eq.~\reff{3}. 
Note that, according to LSI, such a multiplicative
structure is expected in ferromagnetic systems {\it both} for quenches
to $T_c$ and below $T_c$.
The prediction~\reff{lsi0} (with $A_R$, $\alpha$ 
and $\beta$ as fitting parameters) 
has been tested in a variety of different
systems~\cite{Henkel2,Henkel3,Odor,H-wdb,HP-06dis,Henkel,Pleim,Hinrich,noi,noiTRM,noiRd2,Gamba},
with the conclusion that, while it is seemingly successful in a restricted
number of instances, it surely fails in the general case. 
For example, LSI.0 is in agreement with the exact solution
of the large-$N$ (spherical) model~\cite{noininfi,God1,God2} quenched
to $T_{\rm f} \le T_c$, but fails to reproduce all the available analytical 
results~\cite{Lippiello,God1,berthier,EPJ,Mazenko}  
obtained for scalar systems  quenched to $T_{\rm f} < T_c$.
In addition,
for the $d$-dimensional Ising model with Glauber dynamics quenched to
$T_{\rm f} = T_c$, a field-theoretical calculation in the
dimensional expansion $d=4-\epsilon$ 
clearly shows corrections of order $\epsilon^2$ to the
prediction~\reff{lsi0} of LSI.0~\cite{Cala2} for  $d<4$. 
The comparison of LSI.0 with Monte Carlo simulations,
at and below $T_c$, is more controversial since agreement was reported 
in the study of various systems, such as the two- and three-dimensional
Ising model~\cite{Henkel,MHenkel}, the XY model~\cite{Abriet} and the $3$- and
$8$-state Potts
model in two dimensions~\cite{Janke}, 
while deviations from Eq.~\reff{lsi0} were 
detected in the two- and three-dimensional 
Ising model after a quench to the critical 
point in Refs.~\cite{Pleim,noi} and below it in Refs.~\cite{noiTRM,noiRd2}.
In particular, the data presented in Ref.~\cite{Pleim}
for the so-called global intermediate response function 
\be
\label{chiIg-def}
\chi_I^\gl(t,s)=\int _{s/2}^s \rmd t'\, \sum _\r R_r(t,t'),
\ee
and in Ref.~\cite{noi} for $R_{r=0}(t,s)$ 
display deviations from LSI.0 which are actually negligible in $d=4$
but become significant in $d=3$ and even larger in $d=2$,
in qualitative agreement with the expectation based on the
field-theoretical analysis of Ref.~\cite{Cala2}. 

Motivated by these discrepancies,
Henkel {\it et al.}~have reconsidered the original formulation of LSI,
coming up with a new version~\cite{HP-06dis,HenkelLSI1} 
--- referred to as LSI.1 in the following --- which predicts,  in
comparison to  LSI.0, a less constrained scaling form
\be
R(t,s) = A_R s^{-1-\alpha} u^{\beta+ \alpha' - \alpha}(u-1)^{-1-\alpha'}, 
\label{lsi1}
\ee
where $\alpha$ and
$\alpha'$ --- yet undetermined --- are not necessarily equal.
Note, however, that the large-$u$ behavior of Eq.~\reff{lsi1}, being 
actually independent of $\alpha'$, is the same as the one of LSI.0
(see Eq.~\reff{lsi0}) and therefore possible improvements of LSI.1
compared to LSI.0 are restricted to the STSR $u\simeq 1$.
(The space-dependence of $R_r(t,s)$ within
LSI.1 still factorizes according to Eq.~\reff{lsispace}, with $R(t,s)$
given by Eq.~\reff{lsi1}~\cite{HenkelLSI1,H-new}.) 
The introduction of the additional fitting parameter
$\Delta \alpha \equiv \alpha-\alpha'$ 
in Eq.~\reff{lsi1} 
is expected to reduce discrepancies 
in those cases in which
they where observed with $\Delta \alpha = 0$, i.e., when comparing with
LSI.0.
A first apparent confirmation of the predictions 
of LSI.1 came from the comparison with
the {\it numerical} studies of 
(i) the two- and three-dimensional
critical Ising model (Glauber dynamics)~\cite{HenkelLSI1}, 
(ii) the three-dimensional Ising spin glass~\cite{Hen05,HenkelLSI1},
(iii) the non-equilibrium kinetic Ising model in one
dimension~\cite{Odor},
(iv) the contact process~\cite{HenkelLSI1}, and
with the {\it analytical} 
study of the Fredrickson-Andersen
model~\cite{HenkelLSI1}.
However, subsequent analyses of the latter two cases
have revealed clear discrepancies with
LSI.1~\cite{Hinrich,Gamba,MS-07}.

%
In this paper we revise the statement of Ref.~\cite{HenkelLSI1}
according to which the numerical data for 
$\chi_I^\gl(t,s)$ (see Eq.~\reff{chiIg-def}) in the two-dimensional
Ising-Glauber model 
--- though not in agreement with LSI.0~\cite{Pleim} --- do scale
according to LSI.1~\reff{lsi1} with 
$\alpha=a-d/z$, $\beta=\theta$ (see after Eq.~\reff{2}) and
$\Delta\alpha = 0.187(20)$.
More specifically, we shall focus on the fact that the prediction~\reff{lsi1} 
of LSI.1 with $\Delta\alpha\neq 0$ implies a {\it non-stationary}
behavior of the response function in the  STSR: 
\be
R(t=\tau+s,s\gg\tau)=A_R s^{-\Delta \alpha}\,\tau ^{-1-\alpha'}\;,
\label{lsi11}
\ee
which is in qualitative disagreement with what generally 
expected (see Eq.~\reff{1}) also on the basis of
the scaling and RG analyses (Eqs.~\reff{2} and~\reff{aeqa'}).
A conclusive test of LSI.1 for $T_{\rm f} = T_c$ 
can therefore be done by looking at the STSR.
Accordingly, we performed new extensive Monte Carlo simulations
in order to investigate the scaling behavior of the response function
deep in the short-time separation regime, down to $\tau/s \simeq
5\cdot10^{-4}$.
The numerical data presented in Sec.~\ref{MC} 
unambiguously show that $R(t,s)$ becomes time-translational
invariant in the STSR, in agreement with Eq.~\reff{1} and in
stark contrast with the prediction~\reff{lsi11} (with 
$\Delta \alpha\neq 0$) of LSI.1.
Nonetheless, 
the data for $\chi_I^\gl(t,s)$ reproduce the findings of Ref.~\cite{HenkelLSI1}
showing thereby that in passing from $R(t,s)$ to $\chi_I^\gl(t,s)$ 
the space-time integration drastically reduces 
the difference between the actual response function and the prediction
of LSI.1.
We corroborate this statement by studying the scaling behavior of
the {\it local} intermediate response function 
\be
\chi_I^\lo(t,s)=\int _{s/2}^s \rmd t'\, R(t,t'),
\label{chiIl-def}
\ee
which involves only one integration over time and 
which actually deviates from LSI.1 less than in the case of $R(t,s)$
but more than $\chi_I^\gl(t,s)$.
Therefore, in spite of the agreement of 
$\chi_I^\gl(t,s)$ with LSI.1, we conclude that
the prediction of Local Scale Invariance simply provides an 
accurate fitting form for {\it some} quantities but cannot be 
considered to be exact for $T_{\rm f}=T_c$.

\section{Quench to $T=T_c$: Monte Carlo simulations}  
\label{MC}

We consider the Glauber dynamics of the ferromagnetic Ising model 
on a two-dimensional square lattice
with $10^3 \times 10^3$ spins.
The system is initially prepared in an infinite-temperature
configuration and then it is 
quenched to the critical temperature $T_{\rm f}=T_c\simeq 2.269185$.
Temperature is measured  in units of the ferromagnetic coupling $J$ and
time is expressed in Monte Carlo step units (sweeps). 
Each of the data points we shall present in the following is the
result of an average over  $10^4 - 5\cdot 10^4$ different realizations of
the initial condition and of the thermal noise.
No finite-size effects were detected
in the time range accessed during the simulations.

For later convenience, we introduce the (time) integrated response
functions in real and momentum space
\be
\chi (t,[t_1,t_2])\equiv \int_{t_1}^{t_2}\rmd t' R(t,t')
\label{intreal}
\ee
and
\be
\hat \chi_k (t,[t_1,t_2]) \equiv
\int_{t_1}^{t_2}\rmd t' \hat R_k(t,t'),
\label{intmom}
\ee 
respectively, 
where $\hat R_k(t,s) = \sum_{\bf r} R_r(t,s) {\rm e}^{i {\bf r}\cdot{\bf
k}}$ is the Fourier transform of $R_r(t,s)$. 
The various response functions we are interested in, namely 
$R(t,s)$, $\chi_I^\lo(t,s)$, and $\chi_I^\gl(t,s)$ are related to
$\chi (t,[t_1,t_2])$ and $\hat \chi_k (t,[t_1,t_2])$,
which are actually computed without 
applying the external perturbation $h_\x (t)$ via the algorithm
presented in Ref.~\cite{noialg}.
In fact, these response functions can be obtained from
particular correlation functions of the unperturbed system
as~\cite{noialg}:
\be
2 T_c \chi (t,[t_1,t_2])=C(t,t_1)-C(t,t_2)+D(t,[t_1,t_2])
\label{3.1}
\ee
and
\be
2 T_c \hat \chi_k (t,[t_1,t_2])=\hat C_k(t,t_1)-\hat C_k(t,t_2)+
\hat D_k(t,[t_1,t_2]),
\label{3.2}
\ee
where $C(t,t')=\langle \sigma_\x(t)\sigma_\x(t')\rangle$ is the 
spin autocorrelation function, $\langle \dots \rangle$ being the average
over initial conditions and thermal histories, and 
$\hat C_\k(t,t')= \langle \hat \sigma_\k(t) \hat
\sigma_{-\k}(t')\rangle$ is its
momentum counterpart.
The additional correlation functions $D$ and $\hat D$ in
Eqs.~\reff{3.1} and~\reff{3.2} are respectively given by
$D(t,[t_1,t_2])=\langle \sigma_\x(t)B_\x([t_1,t_2])\rangle $
and $\hat D_k(t,[t_1,t_2])= \langle \hat \sigma_\k(t)
\hat B_{-\k}([t_1,t_2]) \rangle$
where $B_\x(t,[t_1,t_2])$ is a 
function of the transition rates and $\hat B_\k(t,[t_1,t_2])$ is its
Fourier transform.
For Glauber dynamics~\cite{Glauber} one has
\be
B_\x([t_1,t_2])= \int _{t_1}^{t_2}\rmd t'\, \left [ \sigma_\x(t') -\tanh
  \left (H_\x^W(t')/T_c\right ) \right ]\;,
\label{3.3}
\ee
where $H_\x^W(t)$ is the Weiss field 
$H_\x^W(t)=\sum_{\langle\x\y \rangle} \sigma_\y(t)$, 
the sum being over sites $\y$ nearest neighbors of $\x$.
Note that the (auto) response $\chi$ and the 
correlation $C$ appearing in Eq.~\reff{3.1}
do not depend on $\x$ due to space homogeneity. In order to reduce the
noise of the resulting numerical data, they have been computed
at each lattice site and then averaged over the whole lattice.

\subsection{Impulse autoresponse function} \label{impulsive}

\begin{figure}[h!]
\centering
\includegraphics[width=12cm]{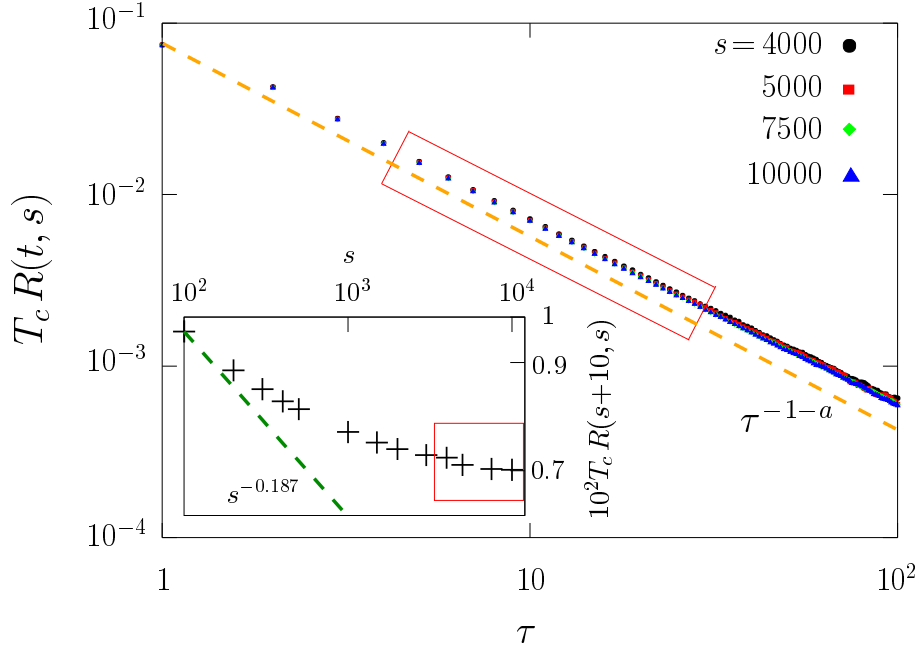}    
\caption{%
Autoresponse function $R(t,s)$ as a function of  $\tau = t - s$
for different values of $s=4\cdot 10^3,5\cdot 10^3,7.5\cdot 10^3$, and
$10^4$.
The dashed line indicates the power-law behavior 
$R(t,s) \sim \tau^{-1-a}$ with $a =0.115$ (see Eq.~\reff{expa}).
The box indicates the points used to fit the value 
of the exponent $a_{\rm MC}$ (see the main text).
%
%
In the inset $R(t,s)$ is plotted as a function of $s$ for fixed 
$\tau =10$.
The dashed line indicates the behavior $\propto s^{-\Delta \alpha}$
predicted by LSI.1, with $\Delta \alpha$ given by Eq.~\reff{delta-a}.
The box contains the points also present in the main figure.}
\label{fig1}
\end{figure}

The impulse autoresponse function $R$ can be formally derived from $\chi
(t,[t_1,t_2])$ (see Eq.~\reff{intreal}) as
$R(t,s)=\lim _{t_2\to s}\chi (t,[s,t_2])/(t_2-s)$. 
Numerically, $R$ is computed as $R(t,s)= (1/\delta )\chi
(t,[s,s+\delta])$, where
the integration over time in $\chi$ has the effect of reducing
the noise-to-signal ratio with the drawback of introducing
a systematic error of order $\delta /s$~\cite{noialg}.  
With an appropriate choice of $s$ and $\delta$, this error can be made
much smaller than the statistical errors and hence neglected. 
In the range of $s$ considered in our simulations ($s>10^2$), 
$\delta = 1$ turns out to be a suitable choice.
The behavior of $R(t,s)$ for $t\gg s$, i.e., in the aging regime, is
quite well established and numerical results for $\chi(t,[0,s])$ (see, e.g.,
Refs.~\cite{God2,Henkel}), $R(t,s)$~\cite{noi,Chat}, and
$\chi_I^\gl(t,s)$~\cite{Pleim} support the validity of the scaling behavior 
predicted within the scaling and
field-theoretical approaches (see Eqs.~\reff{2} and~\reff{aeqa'}):
$R(t\gg s,s) \sim s^{-1-a}(t/s)^\theta$ where the actual 
exponents are compatible with the expected values in $d=2$ (see
Eq.~\reff{expa}): $a=\eta/z \simeq 0.115$
($\eta=1/4$ and $z=
2.1667(5)$~\cite{NB-02})
and $\theta = 0.383(3)$~\cite{GR-95,cg-rev}. 
Accordingly, the comparison 
of the observed behavior of $R(t,s)$ in the aging regime
with LSI.1 fixes $\alpha = a$ and $\beta=\theta$ in Eq.~\reff{lsi1}, leaving
$\Delta\alpha$ as the only undetermined exponent.
In Ref.~\cite{HenkelLSI1} 
\be
\Delta \alpha = 0.187(20) \;
\label{delta-a}
\ee
has been determined as the best parameter value to fit --- actually
with no visible corrections ---
the numerical data of $\chi_I^\gl(t,s)$ 
with the prediction~\reff{lsi1} of LSI.1. 
As pointed out in Sec.~\ref{sec-LSI}, 
if the scaling behavior of
$R(t,s)$ is actually captured by LSI.1 with $\Delta\alpha\neq 0$ then 
in the short-time
separation regime $\tau\gg s \gg \tm$ one will have
$R(t,s) \sim s^{-\Delta\alpha} \tau^{-1-a+\Delta\alpha}$ (see
Eq.~\reff{lsi11}), instead of the generally expected quasi-equilibrium
behavior $R(t,s) \sim \tau^{-1-a}$ 
(see Eq.~\reff{1} and~\reff{3a}). 
In order to test these two qualitatively different predictions we focus on the
behavior of  $T_c R(t,s)$ in the STSR, reported in Fig.~\ref{fig1}.
%
The {\it scaling behavior} is expected to set in for $\tau, s\gg\tm$,
i.e., the actual STSR is restricted to $\tm\ll \tau \ll s$. 
In the inset of Fig.~\ref{fig1} we report, for fixed $\tau=10 \gg
\tm\simeq 1$, the behavior of $T_c R(t,s)$ upon
increasing $s$, eventually probing the STSR. In
particular, according to LSI.1, one expects 
$R(s+10,s)\sim s^{-\Delta\alpha}$, whereas the actual data seemingly
attain a plateau for $s\ge 3\cdot 10^3$ (highlighted by a box in the
inset) setting an upper bound to the asymptotic value of $\Delta\alpha$: 
$\Delta\alpha < 0.029(8)$,
which definitely excludes the value~\reff{delta-a} indicated
in the inset as a dashed line. It seems therefore reasonable to
conclude that $\Delta\alpha=0$ for large enough $s$.
Note, however, that the behavior of 
$R(s+10,s)$ for $s \lesssim 10^3$ could be easily fitted by a power
law with an effective exponent 
$\Delta\alpha_{\rm eff}  \simeq 0.1$
which seemingly increases upon decreasing the largest value of $s$
included in the fit. 
If this observation carries over to the case of $\chi_I^\gl(t,s)$
analyzed in Ref.~\cite{HenkelLSI1},  it might explain
the quite large $\Delta\alpha$~\reff{delta-a} which has been
determined there on the basis of data with $s < 200$.
Repeating the previous analysis for different values of $\tau$ 
we conclude that for $s\ge 3\cdot 10^3$ (highlighted by a box in the inset) 
and $\tau <10^2$ (corresponding to $\tau/s \le 0.032$), $R(\tau+s,s)$
is actually very close to its plateau value in the STSR, although
small corrections of order $\tau/s$ are still present.
The data corresponding to these selected range of $\tau$ and $s$, 
reported in the main plot in
Fig.~\ref{fig1}, display a quite good collapse onto a master curve 
$R(\tau+s,s) \sim \tau^{-1-a_{\rm MC}}$ with $a_{\rm MC}\simeq 0.095(2)$ 
when fitted in the region $5\le \tau \le 30$ (the corresponding points
are highlighted by a box in the main plot of Fig.~\ref{fig1}).
This value is in reasonable
agreement both with $a = 0.115(2)$ and with $a - \Delta\alpha$, 
which are expected on the basis of Eqs.~\reff{1} and~\reff{3a} and of LSI.1 
with $\Delta\alpha < 0.02$, respectively. 
In comparing these values
one has to take into account that our Monte Carlo estimate $a_{\rm
MC}\simeq 0.095(2)$ is seemingly affected by a systematic correction
which reduces $a_{\rm MC}$ compared to its asymptotic value for
$s\rightarrow\infty$.

Although the analysis of the scaling behavior of the response
function in the STSR does not allow us to rule out
the validity of LSI.1, at least it  definitely
indicates that $\Delta\alpha < 0.02$.

\begin{figure}[h!]
    \centering
\includegraphics[width=12cm]{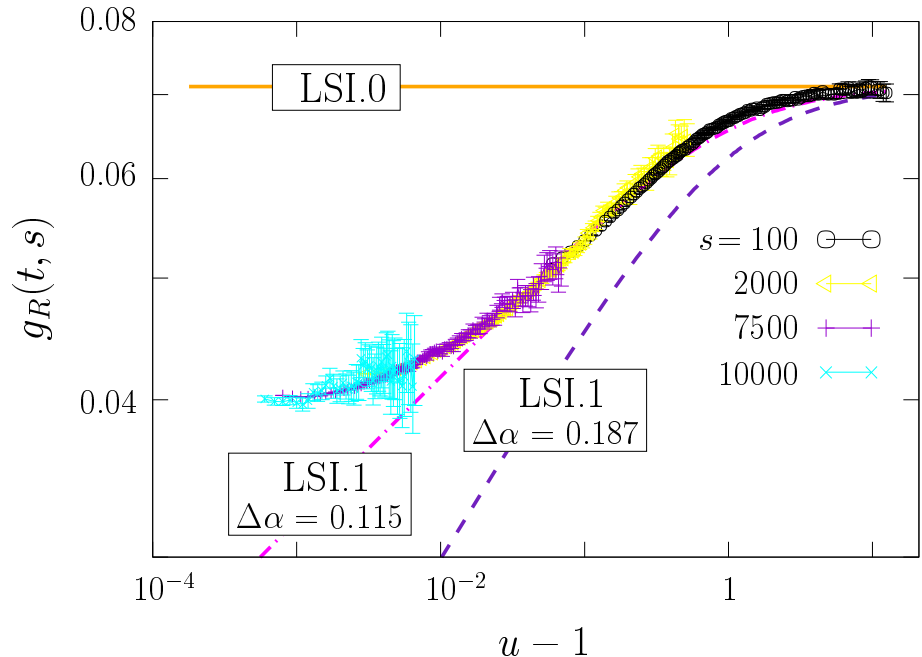}    
\caption{%
Scaling function $g_R(t,s)$ (see Eq.~\reff{3.6}) as a function of 
$u-1$ ($u=t/s>1$) for $s=100$, 2000, 7500, and 10000.
The solid line is the prediction of LSI.0 whereas the dashed and
dash-dotted lines are those of LSI.1 with $\Delta\alpha=0.187$
and $\Delta\alpha=0.115$, respectively. 
In all these cases the values $\theta=0.38$, $a=0.115$,  
and $A_R=0.07$ have been used.}
\label{fig3}
\end{figure}

A more effective comparison between the numerical data and the different
theoretical predictions can be done by considering the quantity
\be  
g_R(t,s) \equiv \tau^{1+a} u^{-\theta} R(t,s),
\label{3.6}
\ee
for arbitrary $t >s$.
According to the general scaling behavior Eq.~\reff{2}, one expects
\be
g_R(t,s) = A_R f_R(u),
\label{3.7}
\ee 
i.e., $g_R(t,s)$ depends only on $u=t/s$ and is such that $\lim_{t\gg
s}g_R(t,s) = A_R$.
In addition, within the RG approach, 
$f_R(u)$ satisfies the condition~\reff{aeqa'},
whereas within LSI.1 $f_R(u)$ is given by 
\be
f^{\rm (LSI.1)}_R(u) = u^{-\Delta \alpha} (u-1)^{\Delta\alpha},
\label{lsi1-fR}
\ee
(compare Eq.~\reff{lsi1} 
with Eq.~\reff{2}) resulting in two different limiting behaviors for
$u\rightarrow 1$:
\be
\lim_{t\rightarrow s} g_R(t,s) =
\begin{cases}
B_R, & \mbox{RG}, \\
A_R(u-1)^{\Delta\alpha}, & \mbox{LSI.1},
\end{cases}
\ee
where $B_R = A_Rf_R(1^+) \neq 0$ (see after Eq.~\reff{3a}).
The qualitative difference between RG and LSI.1 in the STSR translates
into the fact that  for small values of the
abscissa they predict the occurrence of a plateau and a 
power-law behavior with exponent $\Delta\alpha$, respectively, 
when $g_R(t,s)$ is plotted as a function of $u-1$. 
The log-log plot of $g_R(t,s)$ is presented in Fig.~\ref{fig3}
for $s = 10^2$, $2\cdot 10^3$, $7.5\cdot 10^3$, and $10^4$,  
which extend considerably the range of times 
investigated in Refs.~\cite{HenkelLSI1,Pleim}, with $s \le 200$. 
We report only data with $\tau >2$ because,
as discussed in Sec.~\ref{intro}, 
scaling  sets in only for $\tau \gg \tm \simeq 1$.
%
%
For a fixed value of $s$, the statistical uncertainty affecting the
data points in Fig.~\ref{fig3} increases upon increasing  
$u-1$, i.e., $t$, due to the fact that the value of the response
function $R(t,s)\rightarrow 0$ becomes increasingly small and
therefore comparable with statistical fluctuations. As a consequence of
the typical scaling~\reff{genscalag}, this effect is expected to be 
more severe upon increasing $s$, as it is confirmed by comparing 
the data points with $s=10^4$ and $s=10^2$.
The resulting data collapse in Fig.~\reff{fig3}  
is quite good (within the errorbars) 
and confirms that no significant corrections to scaling 
are present.
From the behavior of $g_R(t,s)$ at large $u-1$ one determines the
non-universal constant $A_R$: 
\be
A_R =0.071(2),
\label{AR}
\ee
(obtained by fitting the data with $s=100$ and $u>7$)
whereas in the opposite limit $u\rightarrow 1^+$ one finds (from the
data with $s=7.5\cdot 10^3, 10^4$ and $u-1<1.2\cdot 10^{-3}$):
\be %
B_R = 0.040(1) ,
\label{BR}
\ee
and therefore one estimates
\be
f_R(1^+) = B_R/A_R = 0.56(3).
\label{fRone}
\ee 

\subsubsection{Comparison with the available field-theoretical results}

Note that $g_R(t,s)/A_R = f_R(u)$ characterizes the {\it universal} scaling
behavior of the response function within the Ising universality class
with purely dissipative dynamics. In turn, this fact can be used
in order to determine the associated scaling function in
simplified models, e.g.,
field-theoretical (FT) ones~\cite{FT-Is,Cala2,cg-rev}
belonging to this universality class. 
In particular, in Ref.~\cite{FT-Is} the scaling function for the
spatial Fourier transform of $R_r(t,s)$ (Eq.~\reff{resp}) was determined
up to the first order in $\epsilon = 4 - d >0 $ where $d$ is the
spatial dimensionality of the model.
This result allows the calculation of 
$f_R(u)$ (see Eqs.~\reff{2} and~\reff{3.6}), 
yielding 
\be
f^{\rm (FT)}_R(u) = 1 + O(\epsilon^2),
\label{fRFT}
\ee 
which actually does not reproduce the non-trivial behavior observed in
Fig.~\ref{fig3} and the actual value of $f_R(1^+)$ determined
numerically (see Eq.~\reff{fRone}).
On the other hand, as in the case of the scaling
function of $\sum_{\bf r} R_r(t,s)$ studied analytically in Ref.~\cite{Cala2}, one
expects non-trivial corrections to $f_R^{\rm (FT)}(u)$ at 
$O(\epsilon^2)$ which might
reproduce at least the qualitative features of the numerical
data (such as the fact that $g_R(t,s) \le A_R$).  
The calculation of such corrections, however, is
beyond the scope of the present study.

\subsubsection{Comparison with LSI} 

The mastercurve in Fig.~\ref{fig3} clearly exhibits a plateau at small
$u-1$ which excludes the power-law behavior predicted in this regime 
by LSI.1 
(or, at least, it reduces significantly the upper bound to
$\Delta\alpha$ which results from Fig.~\ref{fig1}~\cite{nota2}).
This conclusion is also confirmed by comparing 
the numerical data for $g_R(t,s)$ 
with the prediction~\reff{3.7}, where $f_R(u)$ is given by
Eq.~\reff{lsi1-fR} and $A_R$ by the fitted value~\reff{AR}. 
In Fig.~\ref{fig3} we present this comparison  
for $\Delta\alpha = 0$ (solid line), corresponding
to LSI.0, $\Delta\alpha=0.187$ (dashed line), corresponding to the
estimate~\reff{delta-a} of Ref.~\cite{HenkelLSI1}, and
$\Delta\alpha=0.115$ (dash-dotted line) ---  which we have 
determined from 
the best fit to the actual data.
Clearly, both  LSI.0 and  LSI.1 with the value
of $\Delta\alpha$ determined in Ref.~\cite{HenkelLSI1} do not capture
correctly the scaling behavior of the response function already for
$u\simeq 6$ and in particular they
overestimate and underestimate the numerical $g_R(t,s)$, respectively.
The choice $\Delta\alpha = 0.115$, however, results in a
good fit of the numerical data down to $u-1 \simeq 2\cdot
10^{-2}$. In spite of that, sizable discrepancies are anyhow observed 
for smaller values of $u-1$ (STSR) and no further improvement is
actually possible by varying $\Delta\alpha$.
In passing we mention that a similar situation is also encountered
when comparing the predictions of LSI.1 with the scaling behavior of
the aging response function within the directed percolation universality
class~\cite{Gamba,Hinrich}.

The conclusion we can draw so far is that the scaling behavior of
the impulse response function $R(t,s)$ {\it does not} agree with the
prediction of LSI.1 even though the latter actually provides a good
fit to the numerical data for $t/s-1 \gtrsim 2\cdot 10^{-2}$. 

\subsection{Integrated local response functions} 
\label{integrated-l}

In this Subsection we discuss the behavior of the local 
integrated response function (see Eq.~\reff{chiIl-def}), which is
given by $\chi_I^\lo(t,s)=\chi (t,[s/2,s])$ in terms of
the quantity~\reff{intreal} measured
in the simulations. This integrated response function has also been
considered in  Refs.~\cite{noiTRM,noiRd2}.

\begin{figure}[h!]
\centering
\includegraphics[width=12cm]{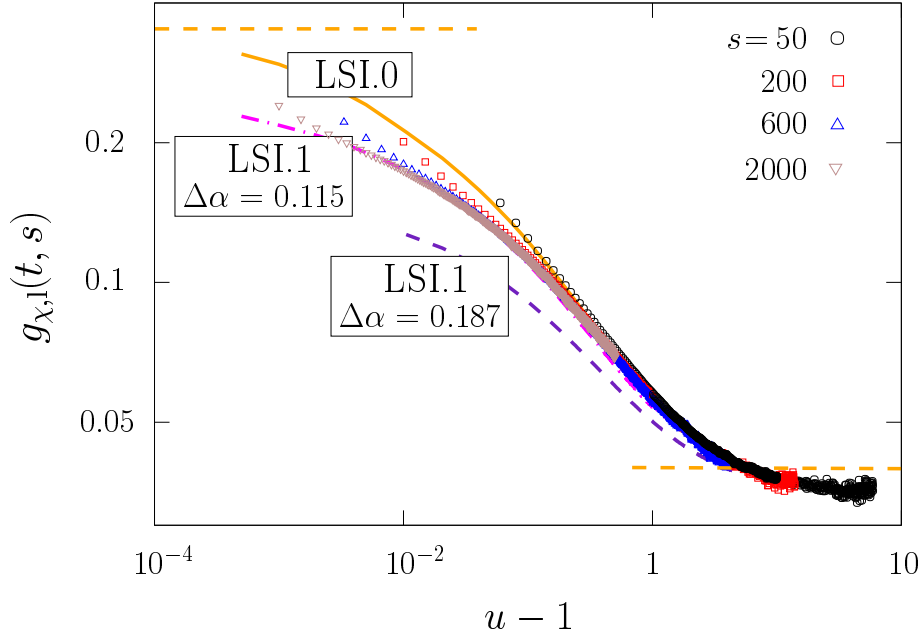}    
\caption{%
Scaling function $g_{\chi,{\rm l}}(t,s)$ (see Eq.~\reff{gchil-def}) 
as a function of $u-1$ 
($u=t/s>1$) for $s=50$, 200, 600, and 2000. 
The solid line is the prediction of LSI.0 whereas the dashed and
dash-dotted lines are those of LSI.1 with $\Delta\alpha=0.187$
and $\Delta\alpha=0.115$, respectively. In all these cases the values
$\theta=0.38$, $a=0.115$, and $A_R=0.07$ have been used.
The horizontal dashed line on the left (right) indicates the limiting
value $B_R/a$ ($A_R \kappa_\theta )$ expected for $u\to 1$ ($u\to
\infty $)
according to the RG and scaling behavior (see Eq.~\reff{limit-gchi})
where $A_R$ and $B_R$ have been determined from the behavior of
$g_R(t,s)$ (see Eqs.~\reff{AR} and~\reff{BR}).}
\label{fig4}
\end{figure}

According to the general scaling behavior of $R(t,s)$ (see
Eq.~\reff{2}), one expects 
\be
\chi_I^\lo(t,s) = A_R  s^{-a} u^{\theta-1} (u-1)^{-a} f_{\chi,{\rm l}}(u)
\label{scal-chiIl}
\ee
where the {\it universal} scaling function $f_\chi^\lo(u)$ is given,
in terms of the scaling function $f_R$, by
\be
f_{\chi,{\rm l}}(u) = (1-u^{-1})^a \int_{1/2}^1 \!\!\rmd x\, x^{-\theta}
\left(1-\frac{x}{u}\right)^{-1-a} f_R(u/x) \;,
\label{chiIl}
\ee
and is such that $f_{\chi,{\rm l}}(u \gg 1) = \kappa_\theta$ with
$\kappa_\theta = (1-2^{-1+\theta})/(1-\theta)$ ($\kappa_\theta \simeq
0.564$ in two dimensions). 
In particular, the prediction of LSI.1 is obtained from
$f_R^{\rm (LSI.1)}(u)$ given in Eq.~\reff{lsi1-fR}:
\be
f^{\rm (LSI.1)}_{\chi,{\rm l}}(u) = \frac{(1-u^{-1})^a}{1-\theta} \left[{}_2F_1(1-\theta,1+a-\Delta\alpha,1/u)
- 2^{\theta-1}{}_2F_1(1-\theta,1+a-\Delta\alpha,1/(2u))\right],
\label{lsi1-fchil}
\ee
where ${}_2F_1$ is the hypergeometric function and the case of LSI.0
is recovered for $\Delta\alpha=0$.
In order to compare the different theoretical predictions we consider
the function
\be
g_{\chi,{\rm l}}(t,s) \equiv  \tau^a u^{1-\theta} \chi_I^\lo(t,s)
\label{gchil-def}
\ee
which, according to Eq.~\reff{scal-chiIl}, is a function of $u=t/s$
\be
g_{\chi,{\rm l}}(t,s) = A_R f_{\chi,{\rm l}}(u)
\label{pred0}
\ee
such that $A_{\chi,{\rm l}} \equiv \lim_{t\gg s} g_{\chi,{\rm l}}(t,s)
= A_R \kappa_\theta$. In the opposite limit, instead,
\be
\lim_{t\rightarrow s} g_{\chi,{\rm l}}(t,s) =
\begin{cases}
B_{\chi,{\rm l}} = B_R/a, & \mbox{RG}, \\
A_R(u-1)^{\Delta\alpha}/(a-\Delta\alpha), & \mbox{LSI.1},
\end{cases}
\label{limit-gchi}
\ee
which is valid under the assumption $a, a-\Delta\alpha >0$ and 
which reflects the different behaviors of
$f_R(u\rightarrow 1^+)$ within RG and LSI.1 (see Eqs.~\reff{aeqa'}
and~\reff{lsi1-fR}, respectively). 
In Fig.~\ref{fig4} we plot $g_{\chi,{\rm l}}(t,s)$  for different
values of $s$, which are slightly larger than those considered in
Ref.~\cite{Pleim,HenkelLSI1} but smaller than the larger values which
Fig.~\ref{fig3} refers to.
%
%
For $s \ge 600$ and $u-1\gtrsim 10^{-2}$ one observes a rather good
collapse onto a single master curve which depends only  on $u$,  in
agreement with the scaling~\reff{scal-chiIl}.
At smaller values of $u-1$ the actual scaling behavior 
is accessed only for increasingly larger values of $s$ such that 
$\tau = t- s > \tm$, i.e., for
$s > \tm/(u-1)$. 
From the large-$u$ behavior of $g_{\chi,{\rm l}}(t,s)$ we determine
the non-universal constant 
\be
A_{\chi,{\rm l}}  = 0.036(3)
\label{Achil}
\ee
in marginal agreement with the expected value 
$A_R \kappa_\theta = 0.040(1)$ resulting from the estimate of $A_R$
given in Eq.~\reff{AR}. 
Note that while $g_R(t,s) \le A_R$, here we find $g_{\chi,{\rm
l}}(t,s) \ge A_{\chi,{\rm l}}$.

\subsubsection{Comparison with the available field-theoretical results:}

The {\it universal} scaling
behavior of $\chi_I^\lo(t,s)$ within the Ising universality class
with purely dissipative dynamics
is characterized by $g_{\chi,{\rm l}}(t,s)/A_{\chi,{\rm l}} = f_{\chi,{\rm
l}}(u)$ 
which can also be studied within
field-theoretical models~\cite{FT-Is,Cala2,cg-rev}.
As a consequence of the fact that $f^{\rm (FT)}_R(u) = 1 + O(\epsilon^2) =
f_R^{\rm (LSI.0)}(u) + O(\epsilon^2)$ 
(see Eq.~\reff{fRFT} and Eq.~\reff{lsi1-fR} with $\Delta\alpha=0$),
one finds, from Eq.~\reff{chiIl}, 
$f^{\rm (FT)}_{\chi,{\rm l}}(u) = f_{\chi,{\rm
l}}^{\rm (LSI.0)}(u) + O(\epsilon^2)$, where $f_{\chi,{\rm
l}}^{\rm (LSI.0)}(u)$ is given by Eq.~\reff{lsi1-fchil} with
$\Delta\alpha=0$. 
The resulting prediction $A_Rf^{\rm (FT)}_{\chi,{\rm l}}(u)$ for
$g_{\chi,{\rm l}}(t,s)$  is shown in Fig.~\ref{fig4} as LSI.0: The
differences with the numerical data are less severe than in the case
of $A_R f_R^{\rm (FT)}(u)$ for $g_R(t,s)$ in Fig.~\ref{fig3}. 
In particular, for $u\rightarrow 1^+$ one has $f^{\rm (FT)}_{\chi,{\rm
l}}(1^+) =  f_{\chi,{\rm
l}}^{\rm (LSI.0)}(1^+) + O(\epsilon^2) =
1/a + O(\epsilon^2)$ (see Eq.~\reff{lsi1-fchil} with $\Delta\alpha=0$
and $u\rightarrow 1^+$)~\cite{footnote}, 
yielding the plateau in the FT prediction
reported in Fig.~\ref{fig4} as a dashed line, 
which overestimates the actual numerical
value. It would be interesting to see whether the corrections of
$O(\epsilon^2)$ to $f^{\rm (FT)}_{\chi,{\rm l}}(u)$ account for
such a difference. 

\subsubsection{Comparison with LSI:} 

As in the case of the impulse response function, the different
qualitative behavior of the LSI.1 and RG predictions should be displayed
for $u-1\rightarrow 0$ according to Eq.~\reff{limit-gchi}, but the
data presented in Fig.~\ref{fig4} are not actually able to detect such
a difference. However, the trend of the data is compatible
with the RG prediction 
$\lim_{t\rightarrow s} g_{\chi,{\rm l}}(t,s) = B_R/a \simeq 0.35$
(see Eq.~\reff{BR}). 
For comparison we 
plot also Eq.~\reff{pred0} with $f_{\chi,{\rm l}}^{\rm (LSI.1)}$ 
predicted by LSI
(see Eq.~\reff{lsi1-fchil}),
$A_R = 0.063$  
(note that this is slightly smaller than the previous estimate
Eq.~\reff{AR})
and for the different values of $\Delta\alpha$ already considered in
Fig.~\ref{fig3}:  $\Delta\alpha = 0$ (solid line), corresponding
to LSI.0, $\Delta\alpha=0.187$ (dashed line), corresponding to the
estimate~\reff{delta-a}
of Ref.~\cite{HenkelLSI1}, and $\Delta\alpha=0.115$ (dash-dotted line).
%
As in the case of $R(t,s)$, both LSI.0 and LSI.1 with
$\Delta\alpha=0.187$ do not provide a good approximation of the actual
data already for $u-1\lesssim 0.1$ and $1$, respectively, whereas LSI.1
with $\Delta\alpha=0.115$ shows visible discrepancies only for
$u-1\lesssim 4\cdot10^{-3}$, when compared to the curve
with  $s = 2000$, 
which is expected to display smaller corrections to scaling.  
The result of the integration over time has been to reduce by almost one order
of magnitude the value
of $u-1$ below which LSI.1 with $\Delta\alpha=0.115$ does not agree with
the numerical data in the scaling regime (from $\simeq 2\cdot 10^{-2}$
for $R(t,s)$ to $\simeq 2\cdot 10^{-3}$ for $\chi_I^\lo(t,s)$). 
By looking solely at Fig.~\ref{fig4} 
one would naturally conclude that the scaling behavior of
$\chi_I^\lo(t,s)$ is actually described by LSI.1 with
$\Delta\alpha=0.115$: The corrections for $u-1\lesssim
4\cdot10^{-3}$ seemingly vanish upon increasing $s$. 

\subsection{Integrated global response functions} \label{integrated-g}

We finally consider the global intermediate response function $\chi_I^\gl(t,s)$
(see Eqs.~\reff{chiIg-def}) 
which was introduced and studied in Ref.~\cite{Pleim} in order to highlight
discrepancies between the numerical data and the prediction of LSI.0
in the case of the  the two- and three-dimensional Ising model with
Glauber dynamics.
The same data presented in Ref.~\cite{Pleim}
have later on been found in agreement 
with the prediction of LSI.1~\cite{HenkelLSI1}.   
In terms of  the quantity~\reff{intmom} measured
in the simulations,  $\chi_I^\gl(t,s)$ is given by
$\chi_I^\gl(t,s)=\hat \chi _{k=0}(t,[s/2,s])$ and it is therefore 
a space integrated, or global, quantity. In fact, $\sum _\r
R_r(t,t')$ is the response function of the total magnetization. 

In this Subsection we focus on the comparison 
between $\chi_I^\gl(t,s)$ and LSI, 
being the one with the corresponding 
field-theoretical predictions~\cite{Cala2}
already discussed in Ref.~\cite{Pleim}.

\begin{figure}[h!]
\centering
\includegraphics[width=12cm]{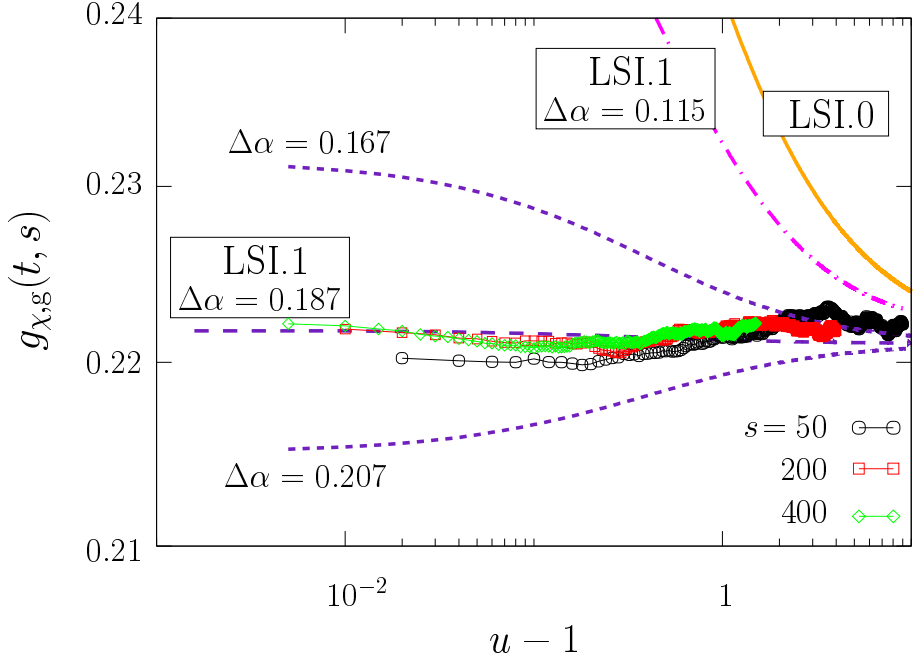}
\caption{%
Scaling function $g_{\chi,{\rm g}}(t,s)$  (see Eq.~\reff{gchig-def})
as a function of $u-1$ ($u=t/s>1$) for $s=50$, 200, and 400.
The solid line is the prediction of LSI.0 whereas the dashed and
dash-dotted lines are those of LSI.1 with $\Delta\alpha=0.187$
and $\Delta\alpha=0.115$, respectively. 
The two additional dashed lines represent the upper and lower bound 
to $\Delta\alpha$ provided in Ref.~\protect{\cite{HenkelLSI1}} and
resulting from the fit of $\chi_I^{\rm (g)}(t,s)$ with the prediction
of LSI.1.
In all these cases
$\theta=0.38$, $a=0.115$, and $\mA_R=0.3895$ 
have been used.}
\label{fig5}
\end{figure}

According to general scaling arguments (see, e.g., Ref.~\cite{Pleim,cg-rev}), the global 
response function $\hat
R_{k=0}(t,s)$ is expected to scale as
\be
\hat R_{k=0}(t,s) = \mA_R
s^{-1-a+d/z}u^\theta(u-1)^{-1-a+d/z}\mF_R(u),
\label{scal-q}
\ee
where the non-universal constant $\mA_R$ is fixed by the
condition $\mF_R(u\gg 1)=1$, and the function $\mF_R(u)$ is {\it
universal}. Within the RG approach one additionally finds that
$\mF_R(1^+)\neq 0$ and {\it finite}, in analogy to Eq.~\reff{aeqa'}.
As a result of the factorization~\reff{lsispace} of the space
dependence of $R_r(t,s)$,  the scaling behavior of the Fourier transform
$\hat R_{k=0}(t,s) = \int \!\rmd{\bf r} R_r(t,s)$ is given by
Eq.~\reff{scal-q} with 
\be
\mF^{\rm(LSI.1)}_R(u) = u^{-\Delta\alpha}(u-1)^{\Delta\alpha},
\label{lsi-q}
\ee 
i.e., it has the same expression as in the case of
$R(t,s)$ but with the formal substitutions $A_R \mapsto {\mathcal
A}_R = A_R\int \!\rmd\x \Phi (\x)$ and $a \mapsto a-d/z$. 
Note that, as in the case of the local response
function, LSI.1 and scaling arguments predicts two qualitatively
different behaviors of $\mF_R(u\rightarrow 1^+)$.
The scaling form of $\chi_I^\gl(t,s)$ is easily calculated on the
basis of Eq.~\reff{scal-q}, yielding
\be
\chi_I^\gl(t,s) = \mA_R t^{-a+d/z} u^{\theta-1} f_{\chi,{\rm g}}(u)
\label{scal-chiIg}
\ee
where
\be
f_{\chi,{\rm g}}(u) = \int_{1/2}^1 \!\!\rmd x\, x^{-\theta}
\left(1-\frac{x}{u}\right)^{-1-a+d/z} \mF_R(u/x),
\label{chiIg}
\ee
such that $f_{\chi,{\rm g}}(u\gg 1) = \kappa_\theta$. In particular
the prediction of LSI.1 is readily found as (see Eqs.~\reff{chiIl} and~\reff{lsi1-fchil})
\be
f_{\chi,{\rm g}}^{\rm (LSI.1)}(u) = 
 \int_{1/2}^1 \!\!\rmd x\, x^{-\theta}
\left(1-\frac{x}{u}\right)^{-1-a+d/z + \Delta\alpha}
= \left[ (1-u^{-1})^{-a} f_{\chi,{\rm
l}}^{\rm (LSI.1)}(u)\right]_{a \mapsto a - d/z} \;.
\label{lsi1-fchig}
\ee
As in the previous cases, in order to compare the numerical data with
the prediction of LSI.1 we consider the quantity
\be
g_{\chi,{\rm g}}(t,s) \equiv t^{a-d/z} u^{1-\theta} \chi_I^\gl(t,s)
\label{gchig-def}
\ee
which, according to Eq.~\reff{scal-chiIg} is a function of $u=t/s$:
\be
g_{\chi,{\rm g}}(t,s) = \mA_R f_{\chi,{\rm g}}(u)
\ee
such that $\mA_{\chi,{\rm g}} \equiv \lim_{t\gg s} g_{\chi,{\rm
g}}(t,s) = \mA_R\kappa_\theta$.
Differently from the previous cases, the behavior of $g_{\chi,{\rm
g}}(t,s)$ for $t\rightarrow s$ is predicted to be
{\it qualitatively the same} both by scaling arguments and LSI.1: 
The integral in
Eq.~\reff{chiIg} is finite for $u\rightarrow 1$, whatever the actual
behavior of $\mF_R(x/u\rightarrow 1^+)$ is, as long as
$\mF_R(u\rightarrow 1^+)\sim (u-1)^{\sigma}$ with $\sigma > a - d/z$,
which is the case here.
In Fig.~\ref{fig5} we plot $g_{\chi,{\rm g}}(t,s)$  for different
values of $s$, which are comparable to those considered in
Ref.~\cite{Pleim,HenkelLSI1} but generally smaller than the values which
Figs.~\ref{fig3} and~\ref{fig4} refer to.
%
%
%
The data sets corresponding to $s = 200$ and 400 collapse to a reasonable
extent, whereas the set corresponding to $s=50$ still departs from the
previous two because of corrections to the scaling behavior. 
For larger values of $u-1$, instead, statistical errors
increase due to a poor statistics. However, our point here is not to
determine the scaling function with high accuracy, but compare its
features to those of the prediction of LSI.1.
As it was already clear from the numerical results presented in Fig.~2(a)
of Ref.~\cite{Pleim} (which Fig.~\ref{fig5} has to be compared 
to~\cite{footnote2}), %
$g_{\chi,{\rm g}}(t,s)$ does not vary too much as a function of $u$,
being almost flat for $0.3 \lesssim u-1 \lesssim 10$ while slightly
increasing upon decreasing $u-1 < 0.3$ (about 2\% in the
interval $0.05\lesssim u-1 \lesssim 0.3$) and decreasing upon
increasing $u-1 > 10$ (by less than 2\%).
Analogous behavior has been observed 
in the three-dimensional Ising model (see
Fig.~2(b) in Ref.~\cite{Pleim}).
From the large-$u$ behavior of $g_{\chi,{\rm g}}(t,s)$ we estimate
the non-universal constant
\be
\mA_{\chi,{\rm g}} = 0.221(1). 
\label{Achig}
\ee %
(Which has actually been obtained on the basis of the data with $s=200$
and $u\gtrsim 3$.)
In Fig.~\ref{fig5} we also plot $\mA_R f_{\chi,{\rm
g}}^{\rm (LSI.1)}(u)$ where $f_{\chi,{\rm g}}^{\rm (LSI.1)}(u)$ in given by Eq.~\reff{chiIg} and $\mA_R = \mA_{\chi,{\rm
g}}/\kappa_\theta$ has been fixed to $0.3895$. 
In particular we report the predictions for $\Delta\alpha = 0$ (solid
line), $\Delta\alpha=0.115$ (dash-dotted line) and
$\Delta\alpha=0.187(20)$ (dashed and dotted lines). 
The prediction of LSI.1 with $\Delta\alpha = 0.187$ provides a
quite good fit of
the actual scaling behavior, in complete agreement with what found
in Ref.~\cite{HenkelLSI1}, where this value of $\Delta\alpha$ was
determined as the best-fit parameter. 
By looking at the integral in the expression of 
$f^{\rm (LSI.1)}_{\chi,{\rm g}}(u)$ 
(see Eq.~\reff{lsi1-fchig}) it becomes clear 
why $f^{\rm (LSI.1)}_{\chi,{\rm g}}(u)$ 
is so accurate in reproducing the almost
constant behavior of the numerical data for $g_{\chi,{\rm g}}(u)$ as
a function of $u$: For $\Delta\alpha = \Delta\alpha_0 \equiv 1 + a -
d/z$, $f^{\rm (LSI.1)}_{\chi,{\rm g}}(u)$ 
becomes constant. In two dimensions one
finds $\Delta\alpha_0 = 0.192(2)$ which is within the interval
determined in Ref.~\cite{HenkelLSI1}.
On the other hand, both the prediction of LSI.1 with $\Delta\alpha=0.115$
--- which was providing the best fit to $R(t,s)$ --- and of LSI.0 are
quite far from the actual numerical data.
In particular, the fact that LSI.1 with $\Delta\alpha=0.115$ well
describes both $R(t,s)$ and $\chi_I^\lo(t,s)$ but not
$\chi_I^\gl(t,s)$ suggests that the factorization of the space 
dependence~\reff{lsispace} might not be a good approximation of the
actual scaling behavior.

On the basis of Figs.~\ref{fig3}, \ref{fig4}, and~\ref{fig5} we can
conclude  that possible integrations  of local quantities over time and over
space strongly affect the agreement  between numerical data and the
corresponding prediction of LSI.1. Accordingly, the exponent
$\Delta\alpha$ which characterizes such a prediction can not be regarded
as an additional model-dependent exponents but simply as a fitting
parameter the optimal value of which depends on the specific quantity
considered. 
Indeed, LSI.1 with $\Delta\alpha =
0.187(20)$ describes correctly the behavior of $\chi_I^\gl(t,s)$ but 
it fails in the case of the related $R(t,s)$. 
Conversely, LSI.1 with $\Delta\alpha = 0.115$ describes very well, for
$t/s - 1> 10^{-2}$, the behavior of both $R(t,s)$ and
$\chi_I^\lo(t,s)$ but it fails in the case of the global response
function $\chi_I^\gl(t,s)$.

\section{Quenches to $T_{\rm f}<T_c$} 
\label{quenchsotto}  

In Sec.~\ref{intro} we have pointed out that 
the prediction~\reff{lsi11} of LSI.1 with $\Delta\alpha\neq 0$ does
not become stationary in the STSR, in contrast to what the response function
actually does. 
Even arguing that the validity of LSI.1 might be restricted only to 
the aging regime does not help for quenches at the
critical point: The multiplicative structure in Eqs.~\reff{3}
and~\reff{3b} ---
a direct consequence of scale invariance --- leaves no room for
LSI.1 to describe $R^{\rm (ag)}$ without spoiling the
quasi-equilibrium behavior of $R(t,s)$ in the STSR.
For $T_{\rm f}<T_c$, however, the crossover between $R^{\rm
(eq)}(\tau;T_{\rm f})$ and $R^{\rm (ag)}(t,s)$ does not occur
multiplicatively but {\it additively}, as we shall discuss shortly. Then, the criticism moved
to LSI.1 for critical quenches does not extend to below $T_c$.

Indeed, in quenches below $T_c$, two-time quantities such as the autoresponse
function naturally split as~\cite{Bouchaud97,Cugliandolo,noiTeff}
\be
R(t,s)=R^{\rm (eq)}(\tau;T_{\rm f})+ R^{\rm (ag)}(t,s).       
\label{2.3}
\ee
This additive structure 
is a direct consequence of the phenomenology of phase-ordering~\cite{Bray},
where a patchwork of compact growing domains, internally in equilibrium,
evolve through defect displacement and/or annihilation.
Fast equilibrium fluctuations inside domains are 
responsible for $R^{\rm (eq)}(\tau;T_{\rm f})$, whereas the slow degrees of
freedom associated to the dynamics of defects give rise to $R^{\rm (ag)}(t,s)$.
One easily verifies that Eq.~\reff{2.3}  renders Eq.~\reff{1} 
in the STSR when $R^{\rm(ag)}$ is given by Eqs.~\reff{2} and~\reff{fRPO} 
(i.e., by LSI.1): 
In this sector, in fact, $u = 1 +
\tau/s \to 1$ and therefore 
$R^{\rm(ag)}(t,s) \sim s^{-\Delta\alpha} \tau^{-1-a+\Delta\alpha}$
which, if $\Delta\alpha>0$, 
becomes negligible for $s\to \infty$ in comparison to
$R^{\rm (eq)}(\tau;T_{\rm f})$. 
In the aging regime, instead, 
the opposite is true: $R^{\rm (eq)}(\tau;T_{\rm f})$ for $u>1$
becomes negligible compared to $R^{\rm (ag)}(t,s)$.
In scalar systems, which we are interested in here, this is due to
the fast (exponential) decay of $R^{\rm (eq)}(\tau; T_{\rm f})$
over a timescale $\tau _{eq}(T_{\rm f})\sim \xi ^{z_g}(T_{\rm f})$, where 
$\xi (T_{\rm f})$ is the correlation length of the equilibrium
state at $T=T_{\rm f}$ with broken symmetry. 
In the aging regime one eventually has
$\tau \gg \tau _{eq}(T_{\rm f})$ and hence $R^{\rm (eq)}(\tau;T_{\rm f})$ 
becomes (exponentially) small compared to $R^{\rm (ag)}(t,s)$ for
$u>1$, yielding $R(t,s) = R^{\rm (ag)}(t,s)$, i.e., 
the response function is very well approximated by $R^{\rm (ag)}(t,s)$. 

Now, as stated in the Introduction, in the analytically solved instances 
of {\it phase ordering} such as the $d=1$ Ising model~\cite{Lippiello,God1} 
quenched to $T_{\rm f}=0$~\cite{footnote1d}
and the time-dependent Ginzburg-Landau model within the 
Gaussian auxiliary field
approximation~\cite{berthier,EPJ,Mazenko}, Eq.~\reff{fRPO} 
holds~\cite{footnoteclass}.
This result has also been verified with high accuracy 
via numerical simulation of the Ising model in
$2,3$ and $4$ dimensions with Glauber dynamics~\cite{noiTRM,noiRd2}, where 
$z_g=2$, and of the one-dimensional 
Ising model with Kawasaki dynamics~\cite{noialg},  
where $z_g=3$. Therefore, in the case of subcritical quenches
the form of $R^{\rm (ag)}$ predicted by LSI.1 
would fit in 
the analytical and numerical results with
\be
\Delta\alpha = 1/z_g.
\label{aneqa'}
\ee
It must be stressed that contrary to the case of
critical quenches, $\Delta\alpha\neq 0$ is not in conflict with the time
translational invariant behavior of the response function 
at short times because of the {\it additive} structure of Eq.~\reff{2.3}.
Note, however, that numerical results for $\chi (t,[0,s])$ and $\hat
\chi_{k=0}(t,[0,s])$ in
the two- and three-dimensional Ising model with Glauber
dynamics~\cite{MHenkel}, quenched to $T_{\rm f} < T_c$, are seemingly
consistent with $\Delta\alpha=0$ 
(see Ref.~\cite{H-wdb} for more details). 
This conclusion might be biased by the fact that these quantities, at variance
with $R(t,s)$ studied in Ref.~\cite{noiRd2}, are affected by quite large
corrections to scaling~\cite{noiTRM}.

\section{Summary and conclusions}   
\label{conclusions}

We have studied in detail, via Monte Carlo simulations, 
the impulse autoresponse function 
$R(t,s)$ [see Eqs.~\reff{resp} and~\reff{autoresp}] and the associated 
(intermediate) integrated {\it local} and {\it global}  response
functions  $\chi_I^{\rm (l)}(t,s)$ [Eq.~\reff{chiIl-def}] and
$\chi_I^{\rm (g)}(t,s)$ [Eq.~\reff{chiIg-def}], respectively, 
of the two-dimensional Ising model with Glauber dynamics, after a
quench from high temperature {\it to the critical point}.

We highlighted the {\it universal} non-equilibrium scaling behaviors
of these quantities and we compared them with the predictions derived
from the theory of Local Scale Invariance in its two recently proposed
versions, referred to as LSI.0 
[Eqs.~\reff{lsi0} and~\reff{lsispace}] and LSI.1
[Eqs.~\reff{lsi1} and~\reff{lsispace}]. 
The scaling form of $R(t,s)$ predicted by LSI.1 depends on the
additional parameter $\Delta\alpha$, compared to the prediction of
LSI.0, which is actually recovered for $\Delta\alpha=0$ (see
Sec.~\ref{intro}).
On the basis of our numerical analysis we conclude that:
%
\begin{itemize}
\item[(i)] $T_c R(t,s)$ becomes --- as generally expected ---
time-translational invariant in the short-time separation regime
which is accessed by increasing $s\gg\tm$ while keeping fixed $\tau =
t-s$ (see Fig.~\ref{fig1}). This behavior is consistent with the
prediction of LSI.0 but not with the one of LSI.1 (with
$\Delta\alpha\neq 0$, see Eq.~\reff{lsi1}). 
\item[(ii)] The scaling behavior of $R(t,s)$ (see Fig.~\ref{fig3} and
Eq.~\reff{3.6}) is correctly captured neither by LSI.0 nor by LSI.1. 
However,  LSI.1 with $\Delta\alpha = 0.115$ yields a good
fit of the actual data only for $t/s-1 > 2\cdot 10^{-2}$. 
Figure~\ref{fig3} --- properly normalized by $A_R$ (see Eq.~\reff{AR}) --- 
provides a very accurate numerical determination of the
{\it universal scaling function} of $R(t,s)$ in a wider range of times
compared to previous numerical studies. 
\item[(iii)] Apparently, the scaling behavior of $\chi_I^{\rm (l)}(t,s)$ (see
Fig.~\ref{fig4} and Eq.~\reff{gchil-def}) is correctly captured by LSI.1 with
$\Delta\alpha = 0.115$. 
The small discrepancies which are visible for $t/s-1 < 4 \cdot
10^{-3}$ can be due either to non-universal correction to scaling (in
which case they should disappear upon increasing $s$) or to the fact that
the actual scaling function of $\chi_I^{\rm (l)}(t,s)$ deviates from
LSI.1 at smaller values of $t/s-1$ (as in the case of $R(t,s)$). The
data at our disposal do not allow one to discriminate between these
two options.
\item[(iv)] In agreement with previous studies~\cite{Pleim}, we find
that the scaling behavior of $\chi_I^{\rm (g)}(t,s)$ (see
Fig.~\ref{fig5} and Eq.~\reff{gchig-def}) is encoded in a scaling
function $g_{\chi,{\rm g}}(t,s)$ which varies less than 4\% in a wide
range of times.
Although the quality of the data presented in Fig.~\ref{fig5} does not
allow an accurate determination of the scaling function, it is
sufficient in order
to conclude that both LSI.0 and LSI.1 with $\Delta\alpha=0.115$ have 
no chance to describe it correctly, whereas --- as noted in
Ref.~\cite{HenkelLSI1} --- LSI.1 with $\Delta\alpha = 0.187(20)$
actually provides a better approximation. 
\end{itemize}
%
The evidences summarized above lead to the conclusion that
the scaling forms predicted analytically 
by LSI.1 for the various quantities
related to $R_r(t,s)$ provide, in some cases, accurate fitting forms
but cannot be considered to be exact, at least for quenches at
the critical point. 
It would be interesting to compare our numerical
determination of the universal scaling function of $R(t,s)$ (see
Fig.~\ref{fig3}) with the predictions of different 
analytical approaches such as, e.g., field-theoretical ones.


\begin{thebibliography}{}
\bibitem{nota2}
Fitting the points in the region
$u-1<10^{-3}$ with $A_{\rm fit} (u-1)^{\Delta\alpha}$
one finds $-0.01 \lesssim \Delta\alpha\lesssim
+0.015$ depending on the value of $s$ which corresponds to the
selected points.
\bibitem{footnote}
%
Within FT, $a = O(\epsilon^2)$ and
therefore $f^{\rm (FT)}_{\chi,{\rm l}}(u)$ up to $O(\epsilon^2)$ 
would be logarithmically divergent for $u\rightarrow 1^+$. On the other
hand, one can improve the estimate of $f^{\rm (FT)}_{\chi,{\rm
l}}(1^+)$ by using in Eq.~\protect{\reff{lsi1-fchil}} the actual
values of the exponents in two dimensions.
\end{thebibliography}
\end{document}